\documentclass[aps, reprint, amsmath, amssymb, superscriptaddress,showpacs]{revtex4-1}
\usepackage[pdftex]{color,graphicx}
\usepackage{amssymb,amsmath}
\usepackage{csquotes}
\usepackage{afterpage}
\renewcommand{\vec}[1]{\boldsymbol{#1}}
\newcommand{\iotabar}{\raisebox{-1pt}{$\mathchar'40$}\mkern-5.43mu\iota}

\begin{document}
\title{3D Tomography of MHD Fluctuations in the H-1NF Heliac}
\author{S. R. Haskey}
\email[]{shaun.haskey@anu.edu.au}
\author{B. D. Blackwell, B. Seiwald,  J. Howard}
\affiliation{Plasma Research Laboratory, Research School of Physics and Engineering, The Australian National University, Canberra, ACT 0200, Australia}
\date{March 2014}

%franz2001soft,piovesan2008tae,jimenez2011alfvenTJ-II,
%%================================================================
\begin{abstract}
A 3D tomographic reconstruction technique is described for inversion of a set of limited-angle high-resolution 2D visible light emission projections (extended in the vertical and toroidal directions) of global MHD eigenmodes in the H-1NF heliac. This paper deals with some of the features and challenges that arise in the application of tomographic imaging systems to toroidal devices, especially limited viewing access and the strong shaping of optimised stellarator/heliotron configurations. The fluctuations are represented as a finite sum of Fourier modes characterised by toroidal and poloidal mode numbers having fixed amplitude and phase in a set of nested cylindrical flux volumes in Boozer space\cite{boozer1980boozer_coords}. The complex amplitude is calculated using iterative tomographic inversion techniques such as ART, SIRT and standard linear least-squares methods. The tomography is applied to synchronous camera images of singly charged carbon impurity ion emission at 514nm obtained at three discrete poloidal viewing orientations \cite{haskey2014synch}. It is shown that the 2D amplitude and phase projections provide high quality reconstructions of the radial structure of the fluctuations that are compact in Boozer space and allow clear determination of the poloidal mode number as well as some degree of toroidal mode number differentiation.

\end{abstract}
\pacs{52.70.Kz, 52.35.-g, 52.70.-m, 52.55.Hc, 52.55.Fa, 42.30.Wb}
\maketitle

\section{Introduction}
\label{sec:intro}
MHD instabilities such as Alfv\'{e}n eigenmodes \cite{heidbrink2008alfven, fasoli2007physics}, sawteeth, and tearing modes can cause detrimental disruptions \cite{Wong1991TAE_TFTR, white1995TFTR, hender2007mhd} and limit achievable plasma parameters. In order to identify and control these modes, accurate measurements of their internal structure are required for comparison with modelling.

\begin{figure}[!ht]
  \begin{center}
    \includegraphics[]{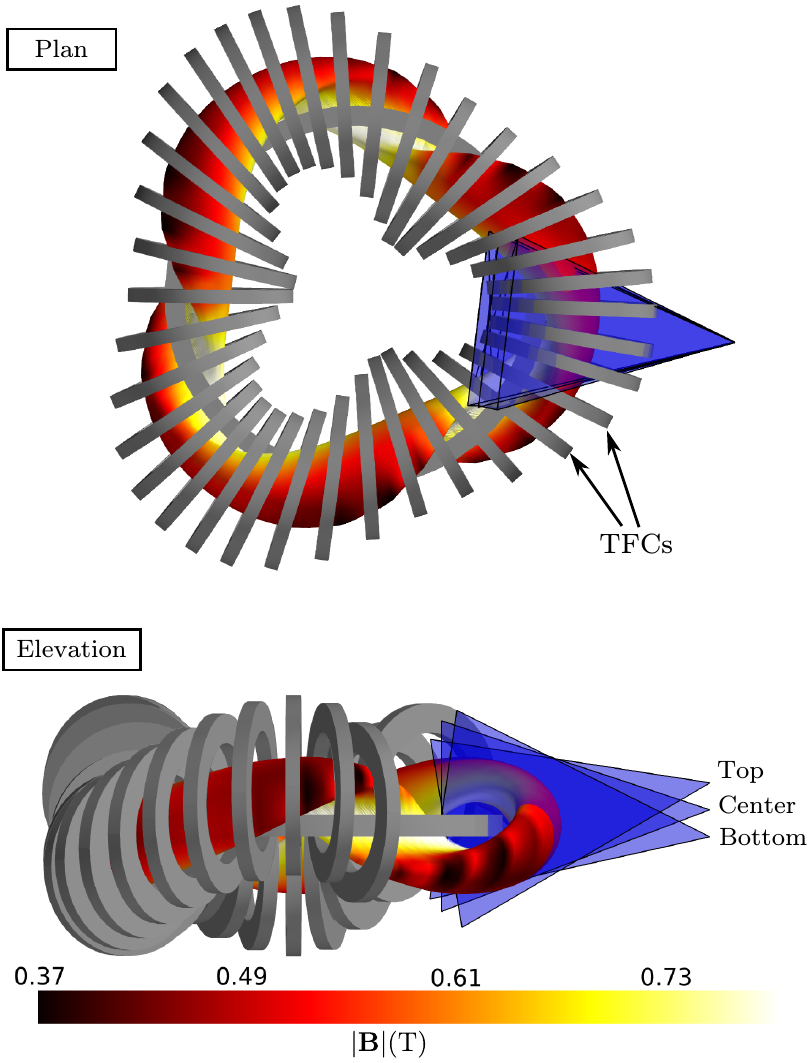}
  \end{center}
  \caption{Details of the three different camera views used to acquire the data on H-1NF. The top plot shows how the toroidal field coils (TFCs) limit the field of view. A typical H-1NF plasma is shown, with the surface colour representing the magnetic field strength.}
  \label{fig:camera_views}
\end{figure}

The structure of MHD fluctuations in the H-1NF heliac \cite{hamberger1990h, jimenez2011alfvenTJ-II} exhibits a systematic dependence on magnetic configuration \cite{blackwell2009configurational, harris2004fluctuations, DaveThesis, pretty2009data}. A longer term goal of the work is to use collisional-radiative models and Bayesian tomographic methods to deduce the structure of the electron density and temperature perturbations from synchronous images of the relative intensities of atomic helium emission line fluctuations for comparison with theoretical predictions \cite{bertram2012ideal,bertram2011reduced}.  As a first step, we here report the tomographic imaging of the mode structure using emission from singly charged carbon impurity ions at 514 nm which is indicative of electron density and temperature fluctuations.

%Constructed using inkscape and raw images from
%/home/srh112/code/python/HMA_analysis/plot_cluster_dimension_plots.py
%/home/srh112/code/python/H1_magnetic/iota_color_plot.py
\begin{figure}[!ht]
  \begin{center}
    \includegraphics[]{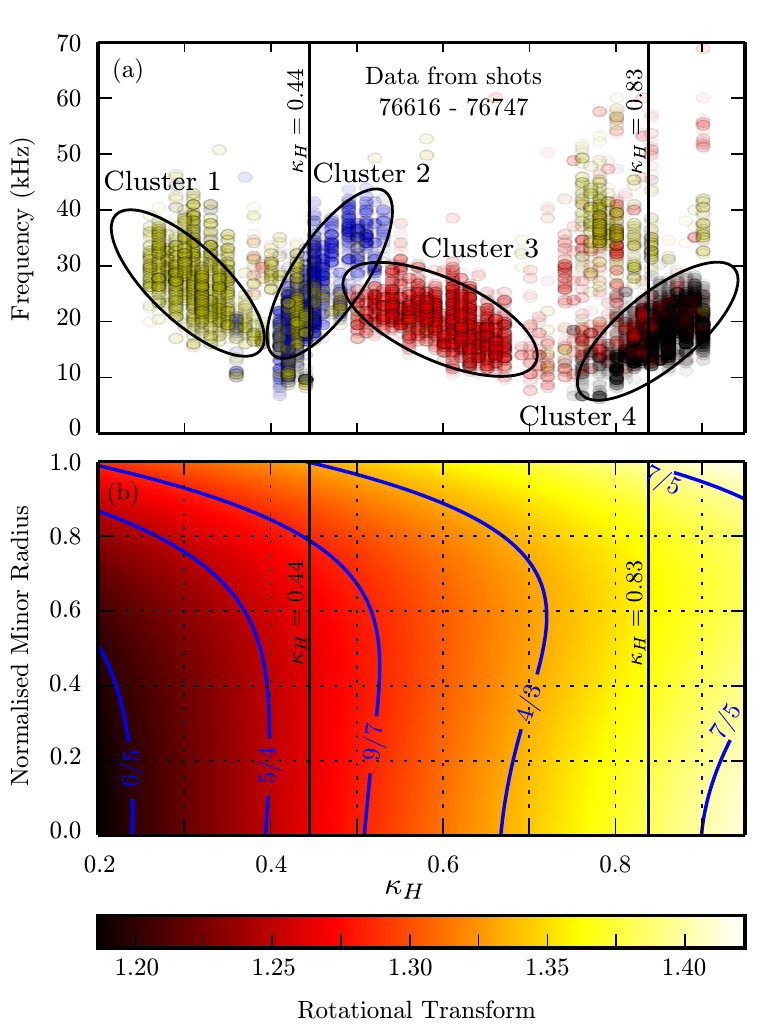}
  \end{center}
  \caption{(a) Fluctuation clusters in H-1NF as a function of frequency and machine configuration parameter $\kappa_H$. Measurements having similar properties are grouped together using a datamining technique. (b) Rotational transform profile of the magnetic field as a function of $\kappa_H$ with the location of some low order rational surfaces shown. See text for more detail.}
  \label{fig:whale_tails}
\end{figure}

Because of the sparseness of available measurement data in plasma physics applications, tomographic inversion methods have generally relied on representations in terms of a small set of continuous, orthogonal basis function such as Fourier-Zernike \cite{granetz1988x, howard_bubble}, and Fourier-Bessel expansions \cite{nagayama1987tomography, franz2001soft}. A recently developed high resolution optical synchronous imaging technique \cite{haskey2014synch}, which uses an intensified camera with gated exposure phase-locked to mode fluctuations allows high resolution imaging of quasi-coherent global plasma eigenmodes with frequencies in the kHz to MHz range. With modern array detectors, these images represent millions of lines of sight which, subject to viewing access, observe a wedge-shaped volume of plasma. Under appropriate assumptions, the measurements can allow high quality tomographic reconstruction of the fluctuation structure to be obtained.

The H-1NF fluctuation structures can generally be represented in terms of a small set of Fourier modes in Boozer coordinates \cite{boozer1980boozer_coords, d1991flux} on a high-resolution discretized radial flux grid.  The linear mapping between projections and the helical plasma modes can then be inverted using standard least squares fitting techniques to obtain the mode amplitudes and phases in the discretized flux regions. The availability of three independent views of the plasma helps determine the helical mode numbers by comparing the best fit error between different candidate mode helicities. Alternatively, the basis set of modes can be imposed {\em a priori} using information from other diagnostics such as magnetic probe arrays \cite{haskey2013hma, hole2009high}.

The tomographic technique described in this paper is general enough to apply to strongly shaped plasmas as well as unconventional viewing geometries in tokamaks. It is well suited to detector-array-based diagnostic systems such as those that measure Bremsstrahlung \cite{van2008tearing}, spectral lines \cite{haskey2014synch} or soft x-ray emission \cite{ohdachi2003high}.  Even with restricted views, the large number of measurements can help constrain poloidal and toroidal mode numbers and can produce detailed radial reconstructions even in the presence of significant noise levels.

This paper is organised as follows. Section \ref{sec:data} provides details of the experimental data used in the tomographic reconstructions and section \ref{sec:tomo} describes the tomographic inversion technique including some of the issues involved in the coordinate transformation of the lines of sight. Section \ref{sec:application_to_h1} provides details of the tomographic inversion, and mode number identification for both odd and even parity eigenmodes in H-1NF.

%=====================================
\section{Data used in the tomographic reconstructions}
\label{sec:data}
The H-1NF heliac is a three field-period helical axis stellarator with major radius $R=1\textnormal{m}$ and average minor radius $\langle r \rangle \approx 0.2\textnormal{m}$ (figure \ref{fig:camera_views}). H-1NF is a flexible machine that allows access to an extensive range of magnetic configurations, making it well-suited to explore the relationship between plasma behaviour and magnetic configuration \cite{harris2004fluctuations}. By varying the ratio of the current in the control helical winding to the current in the toroidal field coils ($\kappa_H$) it is possible to modify the rotational transform profile of the magnetic field as shown in figure \ref{fig:whale_tails} (b). 

As seen in figure \ref{fig:whale_tails} (a), the nature of the magnetic fluctuations in 0.5T, radio-frequency-heated (7 MHz, 40 kW) H/He plasmas \cite{DaveThesis, pretty2009data} in H-1NF is strongly dependent on rotational transform. The fluctuations measured using a magnetic probe array from a 131 shot $\kappa_H$ scan are classified into clusters using datamining techniques\cite{haskey2014EM_VMM}. Each cluster represents a grouping of measured fluctuations that appear similar in spatial structure to the magnetic probe array, and are therefore likely to be due to the same type of fluctuation. Clustering allows the behaviour of certain classes of fluctuations to be studied as a function of plasma parameters.

%multi_view_animation
\begin{figure*}[!ht]
  \begin{center}
    \includegraphics[]{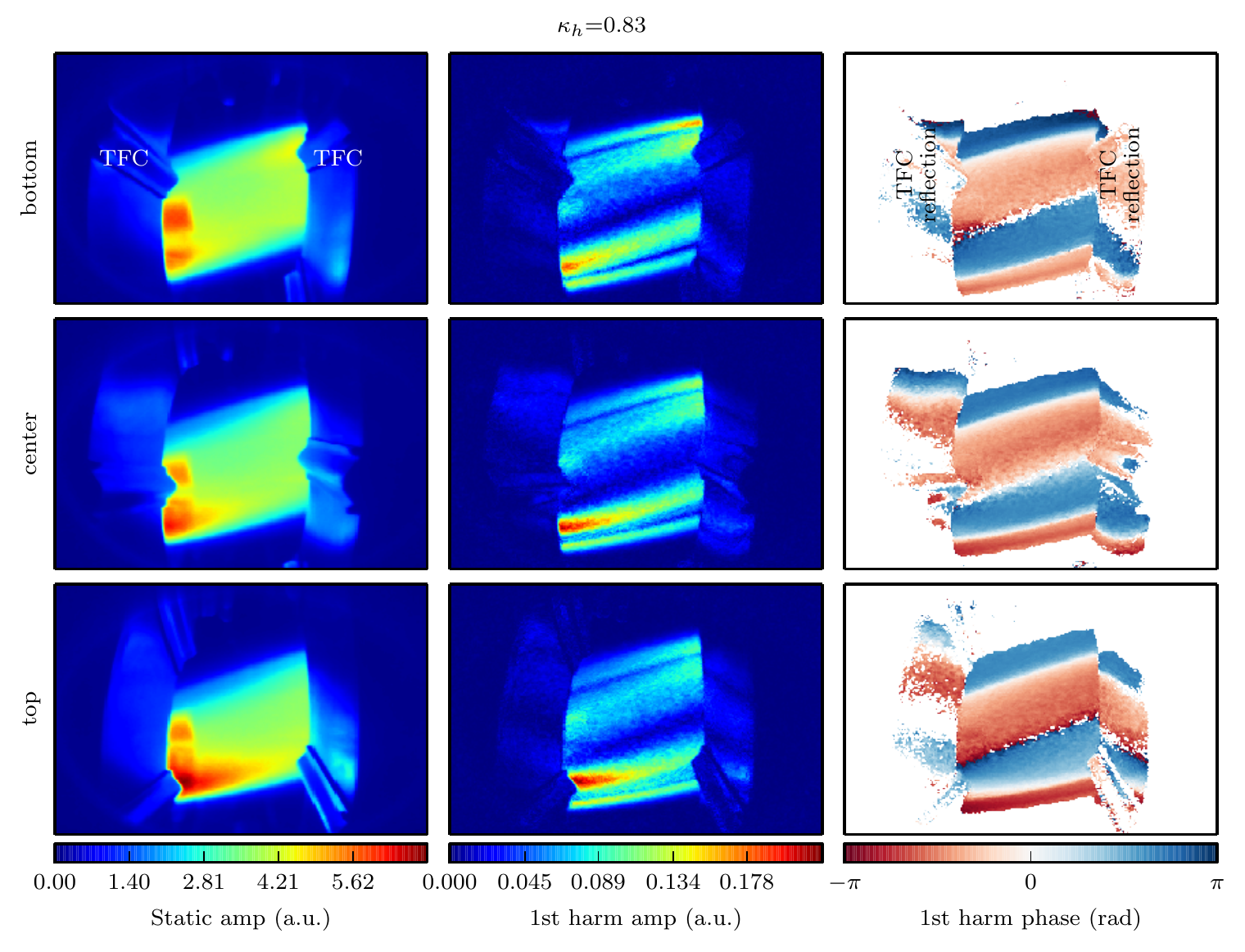}
  \end{center}
  \caption{The static amplitude, 1st harmonic amplitude and 1st harmonic phase of the three separate camera views shown in figure \ref{fig:camera_views}. The fluctuation shown had a frequency of 20kHz, and spontaneously appears in H-1NF discharges with a magnetic configuration of $\kappa_H = 0.83$. The phase images have been thresholded based on the amplitude to remove phase measurements due to noise.}
  \label{fig:camera_images_kh0-83}
\end{figure*}

% \begin{figure}[!ht]
%   \begin{center}
%     \includegraphics[]{./figures/phase_flip_4_3.pdf}
%   \end{center}
%   \caption{}
%   \label{fig:phase_flip}
% \end{figure}

Tomographic reconstructions of two fluctuation structures at $\kappa_H = 0.44$, and $\kappa_H=0.83$ are presented here. These are representative of the behaviour of modes that belong to cluster2 and cluster4 in figure \ref{fig:whale_tails} (a). The waves have frequencies between 20kHz and 25kHz and occur where the zero shear in the  magnetic rotational transform is near the $\iotabar = 5/4$ and $\iotabar = 4/3$ resonances respectively.

The relatively cool electron temperatures in these plasmas means that carbon ion emission at 514nm is radiated from all regions in the plasma, making it suitable for revealing the structural details of the mode, through its dependence on impurity ion density, electron density ($n_e$) and electron temperature ($T_e$). Analysis of photon emissivity using ADAS models \cite{summers2004adas} shows an insensitivity to changes in electron temperature when $T_e>25\mathrm{eV}$, which is typically the case for a large portion of the plasma on H-1NF \cite{ma2012measurements}. Therefore, the tomographic inversions of the intensity fluctuations presented in this paper are closely related to changes in the electron density.

The projection images of the fluctuation structures were obtained using a Princeton Instruments PiMAX 4 intensified gated camera and a synchronous imaging technique \cite{haskey2014synch}. The camera's gated exposure is phase-locked to the mode fluctuation period using a magnetic probe signal \cite{haskey2013hma} as input to a phased-locked loop that produces the camera timing pulses.  The plasma is viewed through the gap between adjacent toroidal field coils at three separate camera inclinations in the poloidal cross-section (figure \ref{fig:camera_views}), the range of viewing angles ($\pm 10^\circ$ wrt the horizontal) being limited by the size of the port window.  The available views provide good coverage of the poloidal cross-section, but are of limited toroidal extent as revealed in figure \ref{fig:camera_views}.% and \ref{fig:boozer_LOS}.

Sixteen camera images equispaced in phase over one cycle of the fluctuation were acquired. These images were decomposed into a Fourier series, allowing the static emission level, fundamental harmonic, and higher harmonics (up to 8) to be isolated. The static emission, and the amplitude and phase projections of the first harmonic for the fluctuation at $\kappa_H=0.83$ are shown in figure \ref{fig:camera_images_kh0-83}. The images show a clear fundamental mode structure.

%=====================================
\section{The tomographic reconstruction technique}
\label{sec:tomo}

\begin{figure}[!ht]
  \begin{center}
    \includegraphics[]{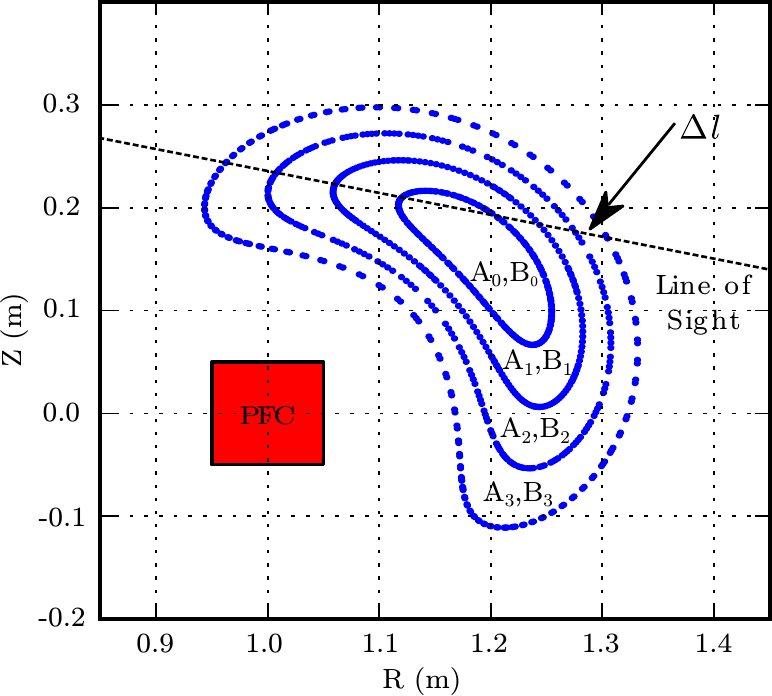}
  \end{center}
  \caption{A typical H-1NF plasma cross-section including the poloidal field coil (PFC). The plasma is split into a series of nested flux regions, where the amplitude and phase $(G_n, \psi_n)$, or real and imaginary components of the mode $(A_n , B_n)$ are constant. The lines of sight are broken up into small length elements, $\Delta l$. While this image is shown in a single poloidal cross section, the method is generalised to lines of sight in any direction.}
  \label{fig:tomo_explanation}
\end{figure}

\subsection{Representation of the perturbation structure}
A scalar perturbation (such as density or light emission) with frequency $f_0$ can be represented as a sum of modes, each having a toroidal and poloidal mode number (n and m):
\begin{equation}
\label{eqn:sum_multiple_modes}
\Phi (s, \theta, \phi, t) = \sum_{n,m} G_{n,m}(s) \cos (n \phi + m \theta + 2 \pi f_0 t + \psi_{n,m}(s))
\end{equation}
where $s=s(x,y,z)$ is the normalised flux (radial variable), and $\phi=\phi (x,y,z)$ and $\theta=\theta (x,y,z)$ are the toroidal and poloidal Boozer coordinates. The amplitude ($G$) and phase ($\psi$) of the modes are assumed to be flux surface quantities in that they depend only on the \enquote{radial} variable $s$.

Initially we treat the case of a perturbation that can be represented by a single toroidal and poloidal mode number:
\begin{align}
\label{eqn:single_mode}
\begin{split}
\Phi (s, \theta, \phi, t) = &G [\cos (n \phi + m \theta +\psi ) \cos (2 \pi f_0 t) - \\
 &\sin (n \phi + m \theta + \psi) \sin (2 \pi f_0 t)]
\end{split}
\end{align}
and return to the sum of multiple modes at the end of this section.

\subsection{Representation of the projections}
The line of sight (LOS) measurement (such as the brightness of a pixel) is given by
\label{sec:tomography}
\begin{equation}
\label{eqn:continuous_line_integral}
P = \int_L \Phi (s, \theta, \phi, t) \mathrm{dl}
\end{equation}
where $L$ is the path along the LOS. Splitting the continuous LOS into $J$ separate intervals of equal length $\Delta l$ gives an approximate discretized version of the line integral
\begin{equation}
P = \sum_{j=0}^J \Phi (s_j, \theta_j, \phi_j, t) \Delta l
\end{equation}
where the subscript $j$ represents the value at the $j$th interval. This is shown in figure \ref{fig:tomo_explanation} where a LOS passes through a series of nested flux regions.

Using a camera synchronized to the fundamental plasma mode frequency $f_0$, it is possible to capture a phase-resolved sequence of projection images \cite{haskey2014synch}.  After Fourier transforming this sequence, the resulting complex pixel brightness for the fundamental component is given by 
\begin{equation}
\begin{split}
\mathcal{P} (f_0) = &\sum_{j=0}^J \left\{\frac{1}{2} G_j \cos (n \phi_j + m \theta_j + \psi_j) \right. \\
& \left. +i \frac{1}{2} G_j \sin (n \phi_j + m \theta_j + \psi_j)\right\} \Delta l
\end{split}
\end{equation}
where $G_j= G(s_j)$ and $\psi_j = \psi(s_j)$ are the amplitudes and phases of the mode at LOS point $j$.

% Extracting the frequency of the mode (the only non zero component):
% \begin{equation}
% \begin{split}
% \label{eqn:probe_output}
% \mathcal{P} (f) =  &\sum_{j=0}^J \frac{1}{2} G_j \cos (n \phi_j + m \theta_j + \psi_j)  [\delta (f - f_0) + \delta (f + f_0)] \Delta l + \\
% &\sum_{j=0}^J \frac{1}{2} G_j \sin (n \phi_j + m \theta_j + \psi_j)  [i \delta (f - f_0) - i \delta (f + f_0)] \Delta l
% \end{split}
% \end{equation}
Separating the real and imaginary components of $\mathcal{P} (f_0)$ and letting $c_j = \cos (n \phi_j + m \theta_j)$, and $\sigma_j = \sin (n \phi_j + m \theta_j)$ gives
\begin{align}
\label{eqn:re_part}
M_R = \mathrm{Re} \{\mathcal{P} (f_0)\} &= \frac{1}{2} \sum_{j=0}^J G_j \cos (n \phi_j + m \theta_j + \psi_j) \Delta l \notag\\
&= \frac{1}{2} \sum_{j=0}^J G_j [c_j \cos(\psi_j) - \sigma_j \sin(\psi_j)] \Delta l
\end{align}
\begin{align}
\label{eqn:im_part}
M_I = \mathrm{Im} \{\mathcal{P} (f_0)\} & = \frac{1}{2} \sum_{j=0}^J G_j \sin (n \phi_j + m \theta_j + \psi_j) \Delta l \notag\\
&= \frac{1}{2} \sum_{j=0}^J G_j [\sigma_j \cos(\psi_j) + c_j \sin(\psi_j)]
\end{align}

%&= \sum_{j=0}^J [c_j A_j + \sigma_j B_j] \Delta l \\
%&= \sum_{j=0}^J [\sigma_j A_j - c_j B_j] \Delta l

Representing the plasma domain as a set of $N$ discrete nested flux surface regions 
$\left\{ s_n| n =0,1, \ldots N \right\}$, and defining $A_n = \frac{1}{2} G_n \cos (\psi_n)$ and $B_n = \frac{1}{2} G_n \sin (\psi_n)$ allows equations \ref{eqn:re_part} and \ref{eqn:im_part} to be written as
\begin{align}
\label{eqn:re_part_flux_reg}
M_R &= \sum_{n=0}^{N-1} \sum_{j\in J_n} [c_j A_{n} - \sigma_j B_{n}] \Delta l\\
\label{eqn:im_part_flux_reg}
M_I &= \sum_{n=0}^{N-1} \sum_{j\in J_n} [\sigma_j A_{n} + c_j B_{n}] \Delta l
\end{align}
where the set $J_n$ selects the segments of the LOS that are in the flux region $n$:
\begin{align}
J_n = \left\{j \in \mathbb{N}| s_n \leq s_j < s_{n+1}\right\}
\end{align}

A set of $X$ pixels corresponding to $X$ lines of sight, provides a set of $2X$ linear equations (equations \ref{eqn:re_part_flux_reg} and \ref{eqn:im_part_flux_reg} for each LOS), which can be represented in matrix form:
\begin{equation}
\label{eqn:tomo_linear_system}
\vec{M} = \vec{S} \vec{T}
\end{equation}
where:
\begin{align}
\vec{M} &= (M_{R,0}, M_{I,0}, M_{R,1}, M_{I,1},...,M_{R,X}, M_{I,X})^T \\
\vec{T} &= (A_0, B_0, A_1, B_1,..., A_{N-1}, B_{N-1})^T
\end{align}

\begin{equation}
\label{eqn:geometry_matrix}
\resizebox{.99\hsize}{!}{$\vec{S} = \left(\begin{array}{ccccccc} C_{0,0} & -S_{0,0} & C_{0,1} & -S_{0,1} & \ldots & C_{0,N-1} & -S_{0,N-1} \\ 
S_{0,0} & C_{0,0} & S_{0,1} & C_{0,1} & \ldots & S_{0,N-1} & C_{0,N-1} \\
C_{1,0} & -S_{1,0} & C_{1,1} & -S_{1,1} & \ldots & C_{1,N-1} & -S_{1,N-1} \\ 
S_{1,0} & C_{1,0} & S_{1,1} & C_{1,1} & \ldots & S_{1,N-1} & C_{1,N-1} \\
\vdots & \vdots & \vdots & \vdots & \ddots & \vdots & \vdots \\
C_{X,0} & -S_{X,0} & C_{X,1} & -S_{X,1} & \ldots & C_{X,N-1} & -S_{X,N-1} \\ 
S_{X,0} & C_{X,0} & S_{X,1} & C_{X,1} & \ldots & S_{X,N-1} & C_{X,N-1} \end{array}\right)$}
\end{equation}
The first and second subscripts on $S_{x,n}$ and $C_{x,n}$ refer to the LOS (or pixel), and flux region respectively. The elements of $\vec{S}$ are calculated as follows:
\begin{align}
S_{x,n} &= \sum_{j\in J_n} \sigma_j \Delta l \\
C_{x,n} &= \sum_{j\in J_n} c_j \Delta l
\end{align}

%/home/srh112/code/python/imax/3D_LOS_interpolation.py
% \begin{figure}[!ht]
%   \begin{center}
%     \includegraphics[]{./figures/boozer_LOS.pdf}
%   \end{center}
%   \caption{Details of the coordinate transformation for the lines of sight of the central camera view for the $\kappa_H = 0.83$ magnetic configuration on H-1NF. Every eighth pixel from a single column is shown.}
%   \label{fig:boozer_LOS}
% \end{figure}

The aim of tomographic inversion is to recover $\vec{T}$ from equation \ref{eqn:tomo_linear_system}. This gives the values of $A_n$ and $B_n$ in each flux region, allowing the amplitude and phase of the associated Fourier component to be recovered as follows:
\begin{align}
\psi_{n} &= \operatorname{arctan2} (B_n, A_n) \\
G_{n} &= \sqrt{(2 B_n)^2 + (2 A_n)^2}
\end{align}
The error in the tomographic reconstruction can be estimated using
\begin{align}
\label{eqn:recon_error}
E = \frac{\sqrt{\langle (\vec{M} - \vec{S} \vec{T})^2\rangle}}{\sqrt{\langle (\vec{M})^2\rangle}}
\end{align}
where $\langle \cdot \rangle$ means an average over all matrix elements, and $(\cdot)^2$ is performed elementwise.

In practice, it is found that the fluctuation structures in H-1NF can be well fitted by summing over a small number of Fourier modes as given in equation \ref{eqn:sum_multiple_modes}.  This type of description is generally suitable for many types of plasma waves in toroidal devices, such as TAE, HAE and EAE's \cite{heidbrink2008alfven}.  The additional modes, $(n_1, m_1),(n_2,m_2)...$, can be accommodated by augmenting the matrices $\vec{S}$ and $\vec{T}$
\begin{eqnarray}
\vec{S} &=& (\vec{S}_{n1,m1}, \vec{S}_{n2,m2}, \vec{S}_{n3,m3},...) \\
\vec{T} &=& (\vec{T}_{n1,m1}, \vec{T}_{n2,m2}, \vec{T}_{n3,m3},...)
\end{eqnarray}
It can also be instructive to calculate the mode emission normalized to the static emission level. The DC emission profile is obtained tomographically by setting $n=0$, $m=0$ and fitting to the dc component of the pixel intensities.

\subsection{Phase variation with flux}
\label{sec:phase_variation}
In the previous sections, the phase ($\psi_{n,m}(s)$) of the mode has been allowed to vary with radius. In the case of wave guide type modes that are represented in an appropriate coordinate system, $\psi_{n,m}(s)$ should not vary with the radial coordinate. It is possible to enforce this by modifying $\vec{S}$ and $\vec{T}$. $\vec{S}$ will have half as many columns and the terms in the odd and even rows are $C_{x,n} \cos (\psi) -S_{x,n} \sin (\psi)$ and $S_{x,n} \cos (\psi) + C_{x,n} \sin (\psi)$ respectively. $\vec{T}= (G_0, G_1,..., G_{N-1})^T$ can then be determined in the same manner as described in section \ref{sec:solution_methods}, although, an optimisation routine which minimises the error (equation \ref{eqn:recon_error}) is required to determine the best $\psi$. 
%In subsequent sections, we use the more general form where the phase is allowed to vary with the radial coordinate

\subsection{Coordinate transformation details}
\label{sec:coordinates}

The Boozer coordinates ($s$, $\theta$, $\phi$) at each interval along the lines of sight must be known in order to calculate $S_{x,n}$ and $C_{x,n}$. For fully three dimensional stellarators this transformation is evaluated numerically using a coordinate system based on the equilibrium plasma magnetic field. The VMEC code \cite{hirshman1983steepest} is used to solve for the plasma equilibrium, and the BOOZ\_XFORM code (part of the STELLOPT package, which includes VMEC) uses this equilibrium to provide the transformation from Boozer co-ordinates to lab coordinates (i.e $x=x(s, \theta, \phi)$, $y=y(s, \theta, \phi)$, $z=z(s, \theta, \phi)$). However, the inverse of this transformation, $s=s(x,y,z)$, $\phi=\phi (x,y,z)$ and $\theta=\theta (x,y,z)$ is required for calculating the projection weight matrix $\vec{S}$.

An efficient method to calculate the inverse transformation uses a dense regular grid in Boozer space that has been transformed to lab coordinates. The forward transformation is relatively fast because the magnetic field is described in terms of a Fourier series.  Using the irregular grid in lab coordinates, a three dimensional interpolation routine is used to approximate the Boozer coordinates at the points along the lines of sight. Due to the periodicity of $\phi$ and $\theta$, a discontinuity exists at the $[0, 2\pi]$ boundary that can cause the interpolation to fail in these regions. This problem can be overcome by performing three separate interpolations with the periodic variable shifted by a fixed amount. The shift moves the discontinuity to different regions of space, thereby allowing the correct values to be selected. Care is also taken in regions where the plasma surface is concave in order to protect against the algorithm calculating interpolates along lines of sight that have exited the plasma. 

Because the coordinates of the LOS are transformed to Boozer coordinates, the technique is general enough to be applied to unconventional viewing geometries in tokamaks and is well suited to strongly shaped devices such as heliotrons/stellarators where simplified viewing geometries are not available. This is of particular importance to fusion relevant devices where there is limited access and port space.

\subsection{Solution and inverse methods}
\label{sec:solution_methods}

The radial profile information, $\vec{T}$ can be recovered from equation \ref{eqn:tomo_linear_system} in a single step using the Moore-Penrose pseudo inverse of $\vec{S}$. This is referred to as the direct solution. If the dimensions of $\vec{S}$, are sufficiently small, this method is very fast, though it scales poorly as the dimensions of $\vec{S}$ increase. The approach can deliver less smooth solutions than alternatives, although this can be overcome using some form of regularization. There are also iterative techniques such as ART \cite{gordon1970algebraic, herman1979image} and SIRT \cite{gilbert1972iterative} that perform more efficiently for large $\vec{S}$. A thorough explanation of these techniques and tomography in general is given in \cite{kak1999principles}.

Each of these methods produced similar results when applied to the data shown in the next section. The direct solution approach has been used for all the results presented in this paper because it was found to be sufficiently smooth, and was substantially faster than ART and SIRT.

\section{Tomography of plasma structures in H-1NF}
\label{sec:application_to_h1}

\subsection{Determining mode numbers}
\label{sec:mode_helicity}
Before performing the tomographic inversion, a set of candidate basis modes must be chosen. This choice can be guided by information from other diagnostics such as magnetic probe arrays or from theoretical modelling. The projections also self-consistently include a great deal of information that constrains the set of allowable helical mode numbers.

%/home/srh112/code/python/imax/3D_LOS_interpolation.py
\begin{figure}[!ht]
  \begin{center}
    \includegraphics[]{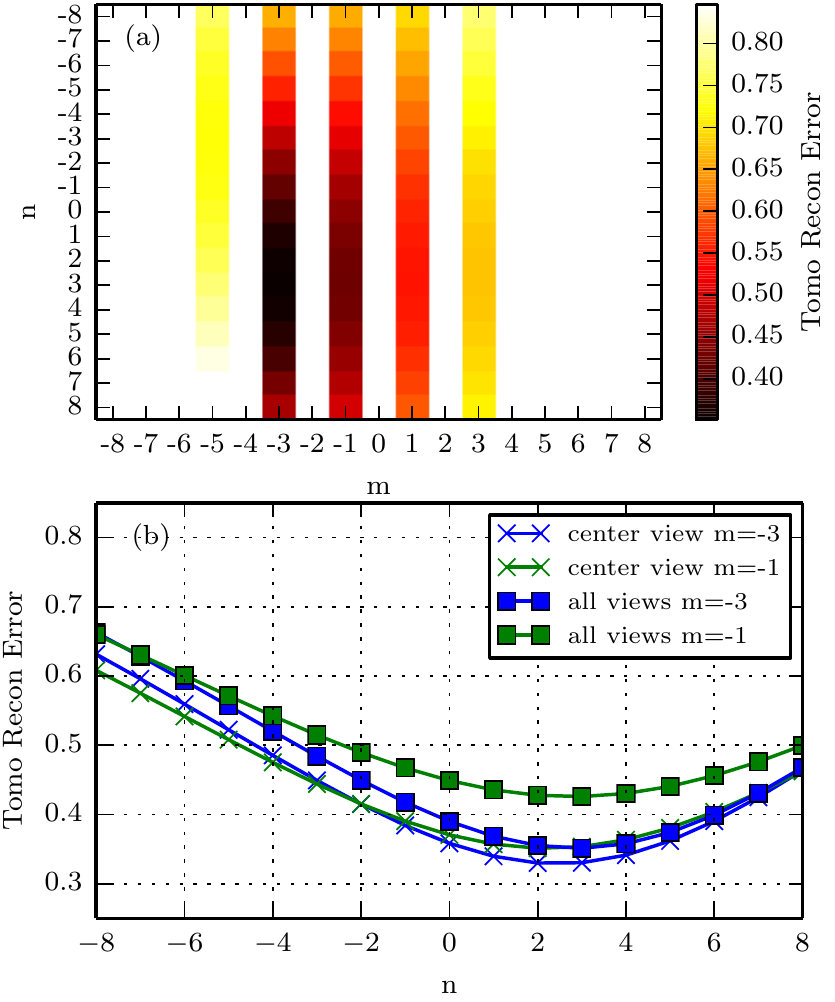}
  \end{center}
  \caption{(top) Tomographic reconstruction error (equation \ref{eqn:recon_error}) using a single mode to fit the data from all three views shown in figure \ref{fig:camera_images_kh0-83}. The technique shows a clear ability to discriminate between poloidal mode numbers using the views on H-1NF. (bottom) The toroidal mode numbers can also be discriminated reasonably well considering the minimal toroidal extent of the views used. This is shown clearly for a variety of $n$ numbers with $m=-3$ and $-1$ using all three views, and using just the center view.}
  \label{fig:single_mode_fits}
\end{figure}

%tomo_sirt.plot_reprojection_comparison(n,m,answer) from 3D_LOS_interpolation.py
\begin{figure*}[!ht]
  \begin{center}
    \includegraphics[]{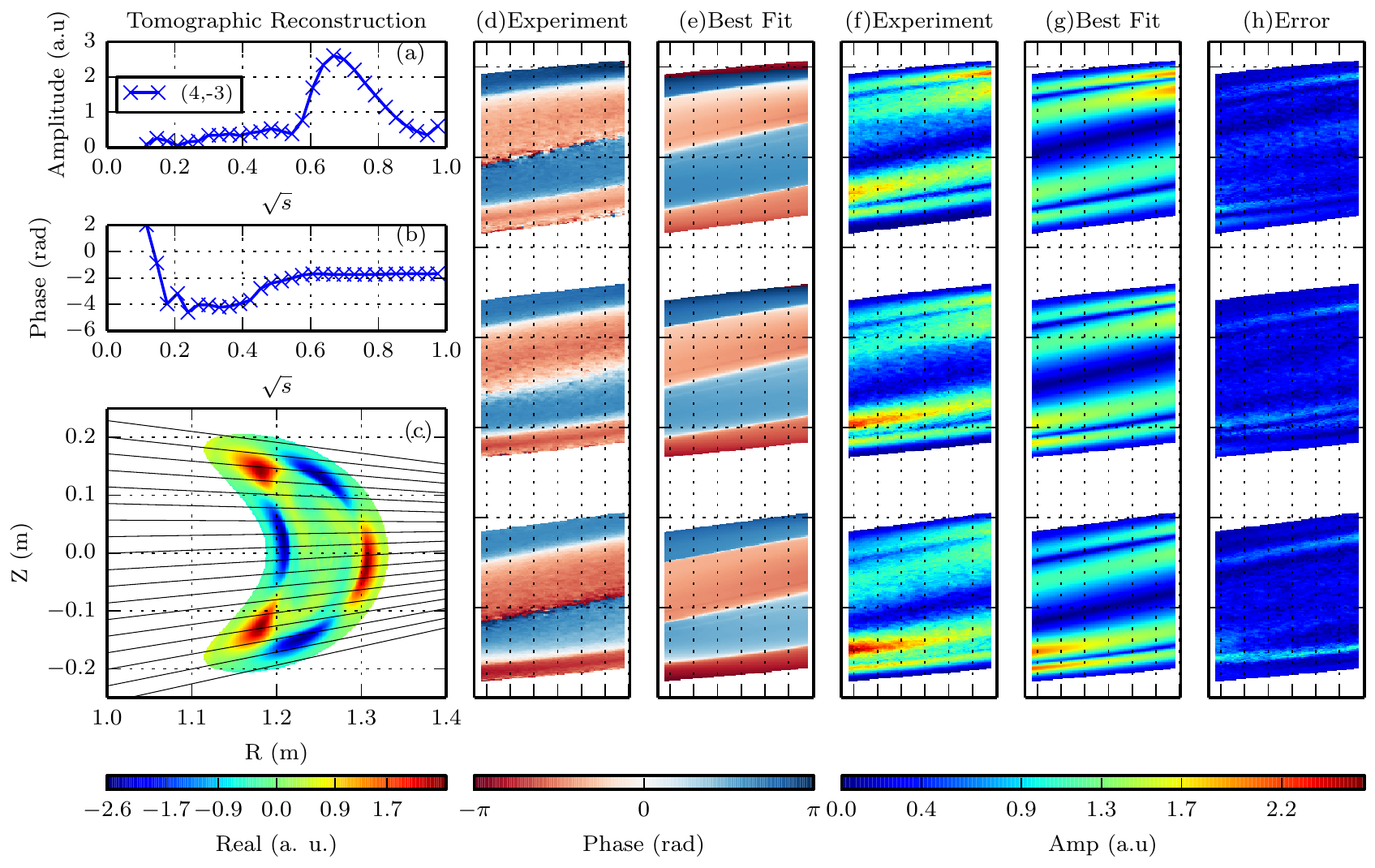}
  \end{center}
  \caption{Details of the tomographic inversion of the camera images shown in figure \ref{fig:camera_images_kh0-83} using a (4,-3) mode for a fluctuation at $\kappa_H = 0.83$. The radial structure of the amplitude and phase are shown in (a) and (b). (c) shows a poloidal cross-section with the real part of the mode plotted and a small subset of LOS from the central camera view. The experimental and best fit phase (d, e), amplitude (f, g) and error in the best fit (h) are also shown. The error is the Euclidean distance between the phasors for each pixel in the experimental and best fit reprojection data. The error is generally low demonstrating that the tomographic inversion is extremely good.}
  \label{fig:reprojection_0.83}
\end{figure*}

%tomo_sirt.plot_reprojection_comparison(n,m,answer) from 3D_LOS_interpolation.py
\begin{figure*}[!ht]
  \begin{center}
    \includegraphics[]{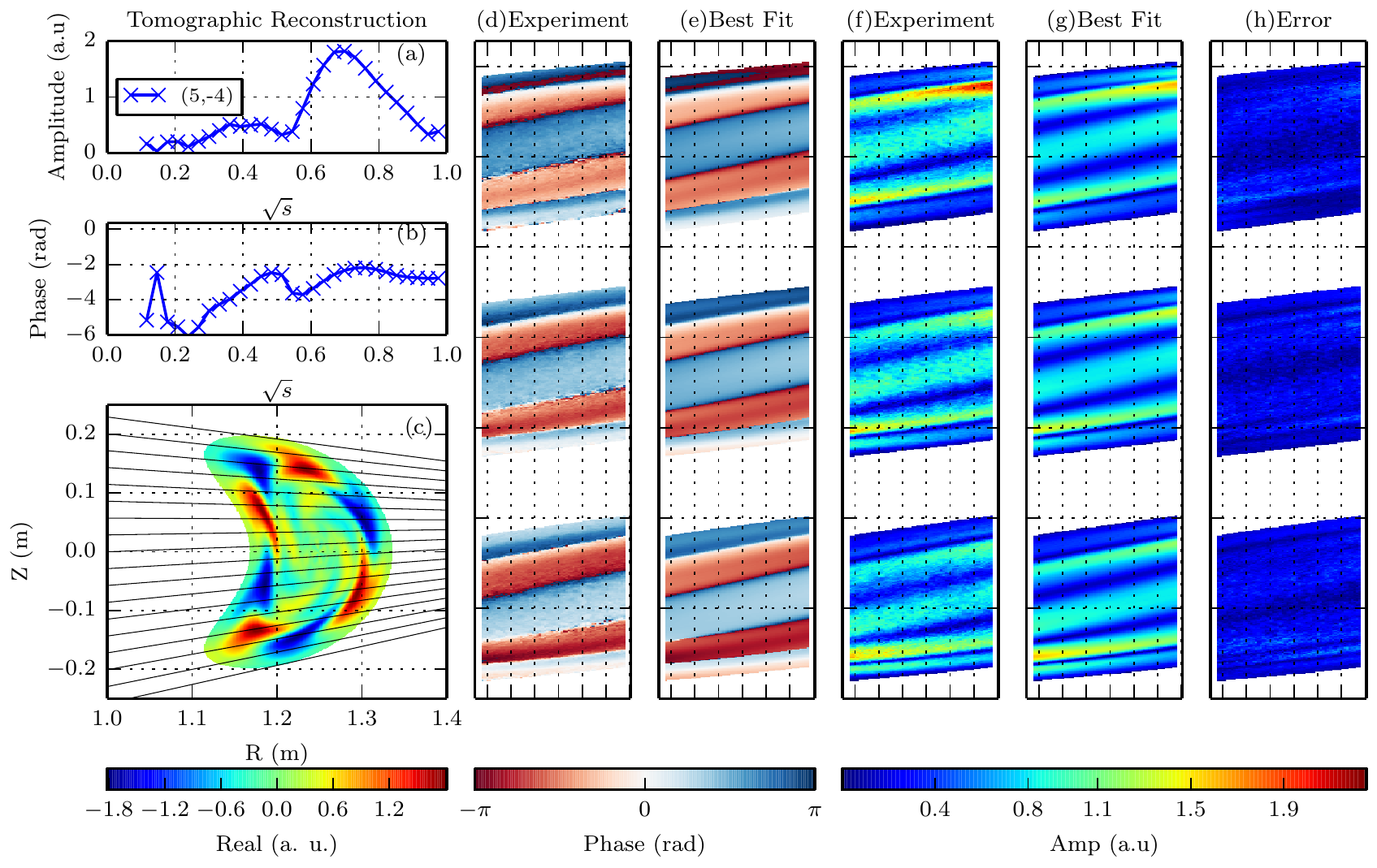}
  \end{center}
  \caption{Details of the tomographic inversion of a fluctuation at $\kappa_H = 0.44$. The subplots are the same as in figure \ref{fig:reprojection_0.83}. The tomographic inversion provides a high quality fit using a (5,-4) mode.}
  \label{fig:reprojection_0.44}
\end{figure*}

Using many trial modes, and calculating the error in the best fit (equation \ref{eqn:recon_error}), it is possible to determine which combination of mode numbers provide the best fit to the data. This is demonstrated in figure \ref{fig:single_mode_fits} (a) where a single mode with $-8 \leq n \leq 8$ and $-8 \leq m \leq 8$ is used in the tomographic inversion of the data from three views shown in figure \ref{fig:camera_images_kh0-83}. Odd poloidal harmonics are clearly favoured by virtue of the phase-reversal across the center of the poloidal projections (see figure \ref{fig:camera_images_kh0-83}). Of the odd poloidal harmonics, $m=-3$ gives the best fit.  This is in accord with analysis of the magnetic probe arrays. The $m=-1$ modes also show an acceptable degree of fit, though higher $m$ structures are clearly incompatible with the number of radial nodes in the projection data.

Figure \ref{fig:single_mode_fits} also reveals that, in spite of the restricted toroidal coverage (see figure \ref{fig:camera_views}), the system shows some limited ability to discriminate between toroidal mode numbers. Including new views at other toroidal locations would provide substantially more toroidal information that would help to identify unambiguously the toroidal mode numbers. For this particular set of images, a toroidal mode number of 3 is favoured, although 1, 2 and 4 also provide acceptable fits. Analysis of the magnetic probe data indicates a toroidal mode number $n=4$.  The slight inconsistency with the camera projections could perhaps be attributed to small errors in the camera registration.

\subsection{Fluctuation radial structure and inversion accuracy}
\label{sec:radial_struct_reproj}
The tomographic inversion of the data in figure \ref{fig:camera_images_kh0-83} using a (4,-3) mode and 30 radial flux regions (evenly spaced in $\sqrt{s}$) is shown in figure \ref{fig:reprojection_0.83}. This mode was chosen because it agreed best with analysis of magnetic probe data and provides a low reconstruction error (figure \ref{fig:single_mode_fits}). The best fit to the projection is accurate across the full toroidal and poloidal extent of all three views. The radial structure of the mode is shown in the left column of figure \ref{fig:reprojection_0.83}. As was discussed in section \ref{sec:data} these radial profiles relate to changes in electron and impurity ion density due to the fluctuation.

The radial profiles are smooth, and show that the mode is radially localised to the outer half of the plasma. The phase of the mode (figure \ref{fig:reprojection_0.83}b) is constant where the amplitude is large ($0.55<\sqrt{s}<0.9$) which is to be expected for a wave guide type mode represented in an appropriate coordinate system. The phase starts to vary considerably for $\sqrt{s} < 0.5$, however, this is not surprising because the phase is not a reliable  quantity in the presence of noise, when the amplitude is small. The amplitude spike at the edge of the reconstruction is an artefact of the inversion process. As the number of radial flux regions is increased up to 120, the spike moves further out and only affects the last one or two flux regions.

For comparison, the tomographic inversion of the even-parity projection data at $\kappa_H = 0.44$ is shown in figure \ref{fig:reprojection_0.44}. Here a single (5,-4) helical mode was fit satisfactorily to the data. As for the previous case, the fit is excellent, and the mode radial structure is smooth and localised beyond mid-radius. In this case, the phase variation with the radial coordinate is larger than in figure \ref{fig:reprojection_0.83}. Performing the tomographic inversions with the phase constrained to be constant (section \ref{sec:phase_variation}) results in an almost identical amplitude profile but with a slightly poorer quality of fit. 

\begin{figure}[!ht]
  \begin{center}
    \includegraphics[]{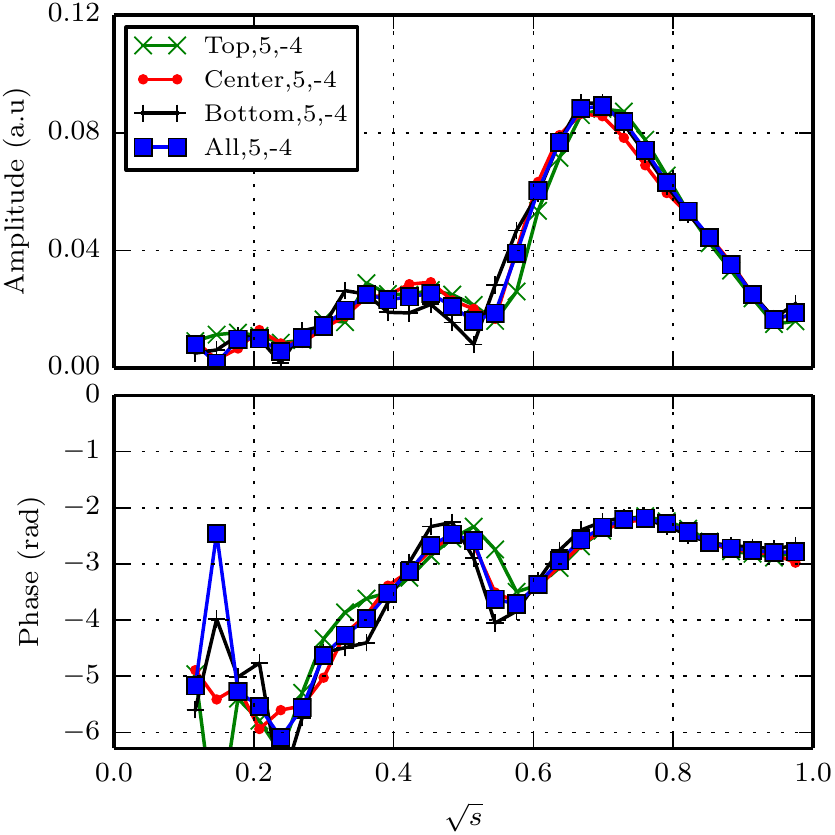}
  \end{center}
  \caption{Tomographic reconstruction of the fluctuation at $\kappa_H=0.44$ using the (5,-4) basis mode for single views, and using all views.}
  \label{fig:view_comparison}
\end{figure}

The tomographic inversions for the $\kappa_H = 0.44$ case (figure \ref{fig:view_comparison}) were also performed using each of the views separately.  In all cases the reconstructions are similar, with the best fit error in the combined case improving by a modest 5$\%$. This suggests that one viewing location can provide sufficient information for the tomographic inversion in the case where a single Fourier mode is dominant.  This could be important to fusion relevant devices where viewing access is limited. The multiple views, however, provide a significant advantage when trying to discriminate between different poloidal mode numbers when {\em a priori} information is unavailable. For example, the difference in the tomographic reconstruction error between the best fit modes with $m=-3$ and $m=-1$ increases threefold when using all three views compared to a single view alone (figure \ref{fig:single_mode_fits} (b)).

%/home/srh112/code/python/imax/3D_LOS_interpolation.py

To improve the fit, it is possible to include additional modes in the basis set used for the inversion. For example using a basis set of (4,-3) and (5,-4) for the $\kappa_H=0.83$ fluctuation improves the fit by 30$\%$. However, a similar reduction in reconstruction error is obtained using (4,-3) and (5,-2) modes, even though there is no reason to expect the latter mode to be present. Improved fits can be obtained using a large number of possible second mode helicities, illustrating that there should be good {\em a priori} reasons for including a second mode in the basis set, and that additional toroidally displaced plasma views are desirable.  In spite of the improvement to the fit, the amplitude of the second even parity mode is relatively small compared to the (4,-3) component.  That the reconstruction of the fluctuation structure is compact (only a single helical mode is required) indicates that the camera view is well registered with respect to the coordinate system and that the Boozer coordinates are a \enquote{natural} basis for the plasma wave.

\subsection{Noise tolerance}
\label{sec:radial_struct_reproj}
\begin{figure}[!ht]
  \begin{center}
    \includegraphics[]{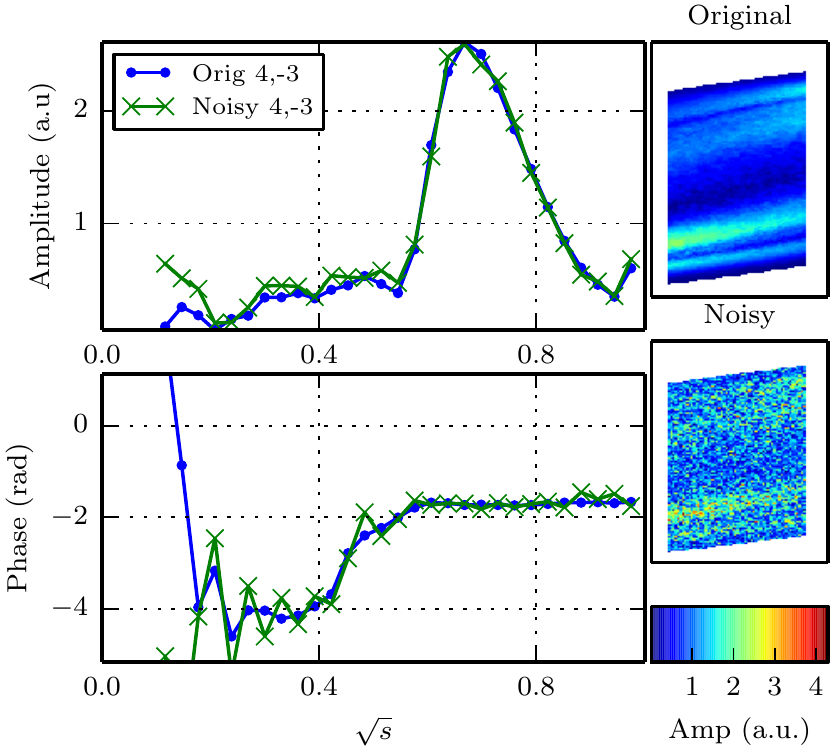}
  \end{center}
  \caption{Details of the tomographic inversion using the original experimental data with and without a significant amount of Gaussian noise added. The reconstructions are similar showing a good tolerance to noise. All three viewing locations were used, but only the images from the center view are shown.}
  \label{fig:noisy_recon}
\end{figure}

Gaussian noise of rms amplitude comparable to the mode fluctuation signal was added to the real and imaginary components of the original experimental data shown in figures \ref{fig:camera_images_kh0-83} and \ref{fig:reprojection_0.83} to test noise sensitivity. The noise added to each of the views was independent, as was the noise added to the real and imaginary parts. Figure \ref{fig:noisy_recon} shows the original amplitude from the center image along with the noisy image. The tomographic inversions are close despite the significant reduction in the signal to noise ratio. This tolerance to noise is due to the large number of measurements available and the fact that only a single harmonic component is fitted. This is particularly important for low amplitude modes and high speed imaging where the reduced exposure time reduces the signal to noise ratio.

%/home/srh112/code/python/imax/3D_LOS_interpolation.py

\section{Conclusion}
\label{sec:conclusion}
A 3D tomographic reconstruction technique that takes advantage of high resolution 2D synchronous imaging and high speed imaging of MHD fluctuations is described. The MHD structures are represented as a sum of Fourier modes in Boozer coordinates with discrete amplitudes and phases in a series of nested magnetic flux regions. This allows the inversion to be represented as a standard linear algebra problem which can be solved using proven tomographic inversion techniques such as ART, SIRT and direct solution. The utility of the technique has been demonstrated on images obtained in 514nm carbon ion impurity light of two different $\approx$ 25kHz MHD structures that appear spontaneously in 0.5T H/He discharges on the H-1NF heliac. 

The technique provides information to aid in toroidal and poloidal mode number identification, produces high quality tomographic reconstructions of the radial structure of the fluctuations and is general enough to apply to strongly shaped plasmas such as those in optimised stellarator/heliotron configurations as well as unconventional viewing geometries in tokamaks. Additionally, the technique is relatively immune to noise, and can be applied to many different camera systems such as those that measure Bremsstrahlung, spectral lines or soft x-ray emission. This combination of uses is particularly important for current and future fusion research devices where there is limited access and port space.

\section{Acknowledgements}
The authors would like to thank the H-1NF team for continued support of experimental operations. This work was supported by the Education Investment Fund under the Super Science Initiative of the Australian Government. SRH wishes to thank AINSE Ltd. for providing financial assistance to enable this work on H-1NF to be conducted.  JH and BB acknowledge support from the Australian Research Council Discovery, grant numbers DP110104833 and DP0666440 respectively.

% \appendix*
% \section{Tomographic inversion details}
% blah blah
% blah

\bibliography{References}

\end{document}